\documentclass[11pt,a4paper]{article}
\pdfoutput=1

\usepackage[T1]{fontenc}
\usepackage{amssymb,amsmath,amsthm,mathtools}
\usepackage{bm}
\usepackage{xcolor}
\usepackage{microtype}
\usepackage{tikz}
\usetikzlibrary{calc}
\usepackage{jheppub}

\definecolor{darkblue}{rgb}{0.0,0.2,0.5}
\hypersetup{
    colorlinks=true,
    linkcolor=darkblue,
    citecolor=darkblue,
    urlcolor=darkblue
}

\newcommand{\Ndat}{N_{\mathrm{dat}}}
\newcommand{\Npar}{N_{\mathrm{par}}}
\newcommand{\thstar}{\hat{\theta}_{\star}}
\newcommand{\thhat}{\hat{\theta}}
\newcommand{\Jstar}{J_{\star}}
\newcommand{\Astar}{A_{\star}}

\newcommand{\bbR}{\mathbb{R}}

\newcommand{\cN}{\mathcal{N}}
\newcommand{\T}{^{\top}}

\begin{document}

\title{Propagating data noise through the fit: the Monte Carlo replica distribution}

\author{Mark N.\ Costantini}
\affiliation{DAMTP, University of Cambridge,\\ Wilberforce Road, Cambridge CB3 0WA, UK}
\emailAdd{mnc33@cam.ac.uk}

\abstract{
The Monte Carlo (MC) replica method quantifies parameter uncertainties in global
fits of parton distribution functions (PDFs) and Standard Model Effective Field
Theory (SMEFT) Wilson coefficients by fitting a model to many noise-perturbed
copies of the data and taking the empirical distribution of the best-fit
parameters as the uncertainty. The method reproduces the Bayesian posterior
exactly only when the model is linear in its parameters, and departs from it
in the nonlinear case. We derive the
leading-order distribution the method produces and compare it with the Laplace
approximation of the Bayesian posterior: the two differ by a single computable
matrix, the residual-weighted Hessian of the model at the best fit, whose sign and
magnitude set the over- or under-estimation of the parameter uncertainties.
This closed-form expression
quantifies when and by how much the MC method departs from Bayesian inference. We
illustrate it on two single-parameter examples solvable in closed form and point to
its evaluation in full PDF and SMEFT fits as a natural next step.
}

\maketitle

\section{Introduction}
\label{sec:intro}

The Monte Carlo (MC) replica method is a procedure for propagating
data uncertainties to fitted parameters via repeated fits on
Gaussian pseudodata. The resulting parameter uncertainties feed
directly into LHC precision measurements and searches for physics
beyond the Standard Model, so the faithfulness of the method is of
practical importance. Introduced in the neural-network PDF programme
of Ref.~\cite{Forte:2002fg}, it has since been widely employed in
parton distribution function (PDF)
fits~\cite{Ball:2008by,Ball:2010de,Ball:2011uy,Ball:2012cx,NNPDF:2014otw,NNPDF:2017mvq,NNPDF:2021njg,Cocuzza:2022hse,Hunt-Smith:2023sdz,Hunt-Smith:2024khs,Ball:2022uon,Courtoy:2022ocu},
in fits of the Standard Model Effective Field Theory (SMEFT) Wilson
coefficients~\cite{Giani:2023gfq,Ethier:2021bye,Ethier:2021ydt,Hartland:2019bjb,Biekotter:2018ohn},
and in simultaneous determinations of
both~\cite{Carrazza:2019sec,Greljo:2021kvv,Iranipour:2022iak,Kassabov:2023hbm,Hammou:2023heg,Costantini:2024xae, Cole:2026eex}.
Much of its success owes to its computational convenience: it replaces
the sampling of a posterior distribution with a set of independent
optimisations, each of which is cheaper than sampling and which can be
run in parallel across replicas.

The same construction, perturbing the data with Gaussian noise and
re-optimising, is not restricted to high-energy physics. An equivalent
construction appears as a method for sampling posterior distributions in
nonlinear inverse problems~\cite{Bardsley:2014} and references therein.
It is closely related to the parametric bootstrap
of statistics~\cite{Efron_2012}, and the same perturb-and-refit idea
underlies ensemble approaches to uncertainty quantification in machine
learning such as randomized prior functions~\cite{Osband:2018}. Across
all of these settings a single fact is well established: the method
reproduces the Bayesian posterior exactly when the model is linear in
its parameters, and departs from it otherwise.

For nonlinear models, this departure can be corrected
numerically: Ref.~\cite{Bardsley:2014} modifies the optimisation so that
the density of the perturbed-and-refitted samples is known in closed
form, and then uses it within a Metropolis--Hastings or importance-sampling
scheme to recover the exact posterior. Such a correction adds a further
layer of computational complexity on top of the bare ensemble of fits.
In this work we characterise
the departure in closed form, deriving a leading-order expression for the
difference between the bare MC replica distribution and the Bayesian
posterior, a single computable matrix. Evaluating it is much cheaper than
the corrective sampling, and tells one, before any corrective step,
whether the bare MC replica distribution already reproduces Bayesian
inference to the required accuracy or whether the correction is needed:
the extra sampling can then be skipped whenever it is unnecessary and run
only when it is not, with the matrix giving in each case the sign and
size of the discrepancy.

In high-energy physics, the parameter distribution implied by the MC
replica procedure was first derived rigorously in
Ref.~\cite{Costantini:2024wby}, which confirmed its agreement with the
Bayesian posterior in the linear case~\cite{DelDebbio:2021whr} and
showed that the two depart in the nonlinear case, with consequences of
practical importance for PDF and SMEFT determinations. More broadly,
the question of how faithfully the MC replica method represents the
posterior is part of a growing interest in Bayesian approaches to PDF
inference~\cite{Candido:2024hjt, Costantini:2025wxp, Costantini:2025agd}.

In this work we give a complementary derivation of the MC replica
distribution that makes the nonlinear departure explicit and computable.
Treating the maximum-likelihood estimator (MLE) as a deterministic map
$d \mapsto \thhat(d)$ from data space to parameter space, we identify
the MC distribution as the pushforward of the Gaussian replica noise
under $\eta \mapsto \thhat(D_{0} + \eta)$. A first-order Taylor
expansion of $\thhat(\cdot)$ around $D_{0}$ reduces this pushforward to
a linear-Gaussian map, whose covariance is fixed by the Jacobian of
$\thhat(\cdot)$ at $D_{0}$; implicit differentiation of the first-order
optimality condition then supplies this Jacobian in closed form,
delivering the leading-order MC distribution. Comparing it with the
Laplace approximation of the Bayesian posterior, we find that the two
differ by a single matrix, the residual-weighted Hessian of the
theory map at the best fit, whose sign and magnitude set the over-
or under-estimation of the parameter uncertainties, and which can be
evaluated directly in a realistic fit.

\paragraph{Outline.}
Section~\ref{sec:mle-map} fixes notation, introduces the MLE as a
data-to-parameter map and computes its Jacobian by implicit
differentiation;
Section~\ref{sec:linear} treats the linear case;
Section~\ref{sec:nonlinear} treats the general nonlinear case and the
comparison with the Bayes--Laplace posterior;
Section~\ref{sec:toys} works out two single-parameter toy examples;
Section~\ref{sec:conclusions} summarises the results and outlines
future directions.

\section{The MLE as a map from data to parameters}
\label{sec:mle-map}

\paragraph{Data, theory, and MLE.}\mbox{}\\
We consider an observed central value $D_{0} \in \bbR^{\Ndat}$, a
sample from the underlying random variable $D \sim \cN(T(\theta), C)$. The theory is defined by a smooth map
$T : \bbR^{\Npar} \to \bbR^{\Ndat}$, $\theta \mapsto T(\theta)$, with Jacobian
\begin{equation}
J(\theta) \;:=\; \frac{\partial T}{\partial \theta}(\theta) \;\in\; \bbR^{\Ndat \times \Npar},
\end{equation}
and Hessian-like third-order tensor $H(\theta) \in \bbR^{\Ndat \times \Npar \times \Npar}$
with components
\begin{equation}
H_{a,ij}(\theta) \;:=\; \frac{\partial^{2} T_{a}}{\partial\theta_{i}\, \partial\theta_{j}}(\theta).
\end{equation}
Dropping the experimental covariance for simplicity, the ordinary
least-squares MLE is\footnote{We note that, in practice, fitting pipelines such as NNPDF do
not literally minimise \eqref{eq:mle-def}: a training/validation split is
used and the stopping criterion is governed by the validation loss rather
than by the global minimum of the training $\chi^{2}$; a more detailed
discussion on this point is given in Appendix~C of
Ref.~\cite{Costantini:2024wby}}
\begin{equation}
\thhat(D_{0}) \;=\; \arg\min_{\theta}\; \tfrac{1}{2}\,\| D_{0} - T(\theta) \|^{2}.
\label{eq:mle-def}
\end{equation}
The general weighted case, in which the Euclidean norm is replaced by the
$C^{-1}$-weighted norm $\| u \|^{2}_{C^{-1}} := u\T C^{-1} u$ for an
experimental covariance $C \in \bbR^{\Ndat \times \Ndat}$, is a
straightforward generalisation: every derivation that follows carries over
by the substitution $\langle u, v \rangle \to u\T C^{-1} v$.
We denote $\thstar := \thhat(D_{0})$, the best fit to the actual data, and we
assume throughout that $\thstar$ is an isolated, non-degenerate local
minimum of \eqref{eq:mle-def}. When the minimum is degenerate, so that
\eqref{eq:mle-def} admits a flat valley of minima rather than an
isolated point, the map $\thhat(\cdot)$ and its Jacobian are not
well-defined. This degeneracy can be lifted by regularising the
(forward) problem, e.g.\ by adding a Tikhonov (ridge) penalty
$\tfrac{\lambda}{2}\|\theta\|^{2}$ to \eqref{eq:mle-def}, equivalent to
a Gaussian prior on $\theta$.

\paragraph{The MLE as a map of the data.}\mbox{}\\
Allowing the data argument to vary in eq.~\eqref{eq:mle-def}, we obtain
the data-to-parameter map
\begin{equation}
\thhat : \bbR^{\Ndat} \to \bbR^{\Npar},
\qquad
\thhat(d) \;:=\; \arg\min_{\theta} \; \tfrac{1}{2}\| d - T(\theta) \|^{2},
\label{eq:phi-def}
\end{equation}
i.e.\ the parameter that best fits the data $d$ in the ordinary
least-squares sense. By construction $\thhat(D_{0}) = \thstar$.

\paragraph{MC replicas as a pushforward of $\thhat(\cdot)$.}\mbox{}\\
The MC replica procedure is the repeated evaluation of $\thhat(\cdot)$ at perturbed
inputs. For $k = 1, \dots, N_{\mathrm{rep}}$, set
\begin{equation}
D_0^{(k)} \;=\; D_{0} + \eta^{(k)}, \qquad \eta^{(k)} \stackrel{\text{i.i.d.}}{\sim} \cN(0, C),
\end{equation}
with $C$ the experimental covariance matrix of the data. By the simplifying
convention adopted in eq.~\eqref{eq:mle-def}, $C$ will be taken to be
$\sigma^{2} I$.
The corresponding replica parameter, written
$\thhat^{(k)}$ as a shorthand for the MLE map evaluated at the $k$-th
pseudodata draw, is
\begin{equation}
\thhat^{(k)} \;:=\; \thhat(D_0^{(k)}) \;=\; \thhat(D_{0} + \eta^{(k)}).
\end{equation}
The MC replica distribution is the law of $\thhat^{(k)}$ across the replica
index $k$, with $D_{0}$ held fixed. It is therefore the pushforward of the
Gaussian noise $\eta \sim \cN(0, C)$ under the map
$\eta \mapsto \thhat(D_{0} + \eta)$: each replica inherits its distribution from
the noise $\eta^{(k)}$, with $D_{0}$ held fixed. All probabilities below are
conditional on $D_{0}$; the only random variable is $\eta^{(k)}$.

\paragraph{Strategy.}\mbox{}\\
For replica noise that is small compared to the data,
the natural step
is to Taylor-expand $\thhat(\cdot)$ around the observed data:
\begin{equation}
\thhat(D_{0} + \eta) \;=\; \thhat(D_{0}) \;+\; \frac{\partial \thhat}{\partial d}(D_0)\, \eta \;+\; O\bigl( \| \eta \|^{2} \bigr).
\label{eq:taylor-Phi}
\end{equation}
Truncating at the linear term reduces the pushforward to a linear map
applied to a Gaussian, which therefore remains Gaussian: at leading
order $\thhat^{(k)} - \thstar = \frac{\partial \thhat}{\partial d}(D_0)\, \eta^{(k)}$
with $\eta^{(k)} \sim \cN(0, C)$, giving the leading-order MC
covariance
\begin{equation}
\Sigma_{\mathrm{MC}}^{\text{leading}} \;=\; \frac{\partial \thhat}{\partial d}(D_0)\, C\, \frac{\partial \thhat}{\partial d}(D_0)\T.
\label{eq:sigma-mc-strategy}
\end{equation}
The problem therefore reduces to three tasks that will be addressed in turn in the next two
subsections: \emph{(a)} verifying that $\thhat(\cdot)$ is a
well-defined function of $d$ on a neighbourhood of $D_{0}$; \emph{(b)}
showing that $\thhat(\cdot)$ is differentiable at $D_{0}$, so
that~\eqref{eq:taylor-Phi} makes sense; and \emph{(c)} computing the
Jacobian $\partial \thhat / \partial d$ at $D_{0}$ in closed form.

\subsection{When the argmin defines a function.}
\paragraph{Linear case.}\mbox{}\\
In the linear case $T(\theta) = J\theta$ the map $\thhat(\cdot)$ is
globally affine in $d$, so tasks~\emph{(a)}--\emph{(c)} of the Strategy
paragraph all trivialise at once and the first-order Taylor
expansion~\eqref{eq:taylor-Phi} is \emph{exact} with no remainder; the
explicit derivation is given in Section~\ref{sec:linear}.

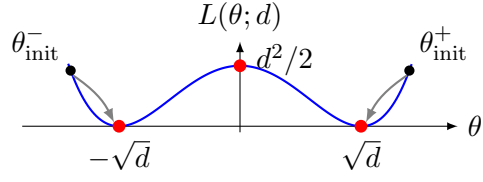
\begin{figure}[t]
\centering
\begin{tikzpicture}[scale=1.6,>=latex]
\draw[->] (-1.8,0) -- (1.8,0) node[right] {$\theta$};
\draw[->] (0,-0.05) -- (0,0.7) node[above] {$L(\theta;d)$};

\draw[blue, thick, smooth, domain=-1.42:1.42, samples=80, variable=\x]
    plot ({\x}, {0.5*(\x*\x - 1)^2});

\fill[red] (1,0) circle (1.5pt);
\fill[red] (-1,0) circle (1.5pt);
\fill[red] (0,0.5) circle (1.5pt);
\node[below] at (1,-0.03) {$\sqrt{d}$};
\node[below] at (-1,-0.03) {$-\sqrt{d}$};
\node[right] at (0.05,0.55) {$d^{2}/2$};

\fill (1.4,0.46) circle (1.2pt);
\node[above right] at (1.4,0.46) {$\theta_{\mathrm{init}}^{+}$};
\draw[->, thick, gray] (1.38,0.42) to[bend right=15] (1.04,0.03);

\fill (-1.4,0.46) circle (1.2pt);
\node[above left] at (-1.4,0.46) {$\theta_{\mathrm{init}}^{-}$};
\draw[->, thick, gray] (-1.38,0.42) to[bend left=15] (-1.04,0.03);
\end{tikzpicture}
\caption{The toy nonlinear loss
$L(\theta; d) = \tfrac{1}{2}(\theta^{2} - d)^{2}$ for $T(\theta) = \theta^{2}$
at $d > 0$. Two equally good fits at $\theta = \pm\sqrt{d}$ are
separated by a local maximum at $\theta = 0$. The initialisation
$\theta_{\mathrm{init}}$ selects which basin gradient descent falls
into and hence which of the two minima the fit returns.}
\label{fig:double-well}
\end{figure}

\paragraph{Non-linear case.}\mbox{}\\
In the nonlinear case the argmin need not be unique. A simple illustration is the scalar example
$T(\theta) = \theta^{2}$ with $\Npar = \Ndat = 1$. For any $d > 0$ the
loss
\begin{equation}
L(\theta; d) \;=\; \tfrac{1}{2}\bigl( \theta^{2} - d \bigr)^{2}
\end{equation}
is a symmetric double-well in $\theta$, with two equal global minima
at $\theta = \pm\sqrt{d}$ (where $L = 0$) separated by a local maximum
at $\theta = 0$ (where $L = d^{2}/2$); see Fig.~\ref{fig:double-well}.
The model is not injective, $T(\theta) = T(-\theta)$, so the same
prediction is produced by two different parameter values, and the
argmin in~\eqref{eq:phi-def} is genuinely two-valued. Which of the two
minima a fit returns is determined entirely by the optimiser's starting
point: gradient descent initialised at $\theta_{\mathrm{init}} > 0$
crosses the right basin and converges to $\theta = +\sqrt{d}$, while
$\theta_{\mathrm{init}} < 0$ leads to $-\sqrt{d}$, with the local
maximum at $\theta = 0$ acting as the basin boundary. The same
phenomenon reappears in the higher-dimensional fits relevant in
practice where multiple local minima of the
loss generically coexist.

\paragraph{The initialisation selects a local minimum.}\mbox{}\\
What an actual fit returns is therefore not the global argmin
in~\eqref{eq:phi-def} but the specific local minimum the optimiser
converges to from its starting point. We accordingly \emph{refine} the
definition~\eqref{eq:phi-def} by promoting the initialisation to an
explicit argument: from now on
\begin{equation}
\thhat(d; \theta_{\mathrm{init}}) \;:=\; \text{the local minimum of \(\tfrac{1}{2}\| d - T(\theta) \|^{2}\) reached from } \theta_{\mathrm{init}},
\label{eq:phi-def-refined}
\end{equation}
implicitly characterised by the first-order condition (FOC), obtained
by setting the gradient of the loss
$\tfrac{1}{2}\| d - T(\theta) \|^{2}$ with respect to $\theta$ to zero,
\begin{equation}
F(\theta, d) \;:=\; J(\theta)\T \bigl( T(\theta) - d \bigr) \;=\; 0,
\label{eq:foc}
\end{equation}
together with the choice of basin selected by
$\theta_{\mathrm{init}}$.\footnote{The basin reached may also depend on
the optimisation procedure itself: different deterministic algorithms
can land in different basins from the same $\theta_{\mathrm{init}}$,
and stochastic optimisers make $\thhat(\cdot)$ a random variable. We
ignore these complications throughout.}
The initialisation plays a passive but essential role: it labels which
of the (potentially many) local minima of the loss the function $\thhat(\cdot)$
returns. In the linear case the dependence on $\theta_{\mathrm{init}}$
is trivially absent, and~\eqref{eq:phi-def-refined} reduces
to~\eqref{eq:phi-def}.

In a Monte Carlo replica simulation one fixes a single initialisation
$\theta_{\mathrm{init}}$ and reuses it across all replica fits. With
this convention, the MC replica distribution is the pushforward of
$\cN(0, C)$ under $\eta \mapsto \thhat(D_{0} + \eta; \theta_{\mathrm{init}})$,
and $\thstar := \thhat(D_{0}; \theta_{\mathrm{init}})$ is the resulting best
fit to the actual data $D_{0}$. Differentiability at $D_{0}$, and the
closed-form Jacobian, are the subject of the next subsection.

\subsection{Computing the Jacobian.}
\paragraph{Implicit differentiation of the FOC.}\mbox{}\\
We derive the Jacobian $\partial \thhat / \partial d$ at $D_{0}$ by
implicit differentiation of the FOC~\eqref{eq:foc}. By construction,
$\thhat(d; \theta_{\mathrm{init}})$ is a local minimum of the loss and
therefore satisfies 
$F\bigl( \thhat(d; \theta_{\mathrm{init}}), d \bigr) \equiv 0$.
Taking the total derivative of this identity with respect to $d$ and
evaluating at $(\thstar, D_{0})$, the chain rule gives
\begin{equation}
\partial_{\theta} F \,\cdot\, \frac{\partial \thhat}{\partial d}
\;+\; \partial_{d} F \;=\; 0.
\label{eq:implicit-diff}
\end{equation}
The data derivative is immediate from
$F(\theta, d) = J(\theta)\T(T(\theta) - d)$:
$\partial_{d} F = -\Jstar\T$, with $\Jstar := J(\thstar)$. The parameter derivative
$\partial_{\theta} F$ is the Hessian of the loss, and is also straightforward to 
compute as we show in the next paragraph.

\paragraph{The Hessian of the loss.}\mbox{}\\
Since $F$ is itself the gradient of the OLS objective
\begin{equation}
L(\theta; d) \;:=\; \tfrac{1}{2}\,\| d - T(\theta) \|^{2}
\;=\; \tfrac{1}{2}\sum_{a=1}^{\Ndat} \bigl( d_{a} - T_{a}(\theta) \bigr)^{2},
\end{equation}
($\partial L / \partial \theta_{i} = F_{i}$), $\partial_{\theta} F$ is
just the Hessian $\nabla_{\theta}^{2} L$ of the loss. To obtain it
explicitly, we differentiate
$F_{i}(\theta, d) = \sum_{a} J_{a,i}(\theta)\,(T_{a}(\theta) - d_{a})$ in
$\theta_{j}$ and use the product rule. The two factors $J_{a,i}(\theta)$ and
$T_{a}(\theta) - d_{a}$ each contribute one term:
\begin{equation}
\frac{\partial F_{i}}{\partial \theta_{j}}(\theta, d)
\;=\; \sum_{a=1}^{\Ndat} \biggl[ \underbrace{\frac{\partial J_{a,i}}{\partial \theta_{j}}(\theta)}_{= \, H_{a,ij}(\theta)} \bigl( T_{a}(\theta) - d_{a} \bigr) \;+\; J_{a,i}(\theta)\, \underbrace{\frac{\partial T_{a}}{\partial \theta_{j}}(\theta)}_{= \, J_{a,j}(\theta)} \biggr],
\end{equation}
which in matrix form is
\begin{equation}
A(\theta, d) \;:=\; \partial_{\theta} F(\theta, d)
\;=\; J(\theta)\T J(\theta) \;+\; \sum_{a=1}^{\Ndat} \bigl( T_{a}(\theta) - d_{a} \bigr)\, H_{a,\cdot\cdot}(\theta).
\label{eq:hessian}
\end{equation}
The first term of~\eqref{eq:hessian}, also called the Gauss--Newton matrix, is the Hessian 
one would obtain by linearising $T$ at $\theta$. In particular, this term can be seen
as a good approximation of the Hessian of a nonlinear least-squares problem whenever the residuals are small.
The second term is the \emph{residual-weighted Hessian} of $T$ and encodes the curvature of $T$ in data space,
weighted by the residual $T_{a}(\theta) - d_{a}$. This second term, in particular, vanishes whenever the
model fits the data perfectly or $T$ is linear in $\theta$.
Evaluated at $(\thstar, D_{0})$ it gives
\begin{equation}
\Astar \;:=\; A(\thstar, D_{0})
\;=\; \Jstar\T \Jstar \;+\; \sum_{a=1}^{\Ndat} r_{a}\, H_{a,\cdot\cdot}(\thstar),
\qquad r_{a} \;:=\; T_{a}(\thstar) - D_{0,a}.
\label{eq:Astar-decomp}
\end{equation}

\paragraph{Differentiability of $\thhat(\cdot; \theta_{\mathrm{init}})$.}\mbox{}\\
The map $\thhat(\cdot; \theta_{\mathrm{init}})$ defined in
Section~\ref{sec:mle-map} is in fact a $C^{1}$ function of $d$ on a
neighbourhood of $D_{0}$, by the implicit function theorem applied to the
FOC residual $F$ at $(\thstar, D_{0})$:

\medskip

{\sloppy
\noindent\textbf{Theorem (IFT).}
\emph{Suppose $F(\theta, d)$ is continuously differentiable near
$(\thstar, D_{0})$ with $F(\thstar, D_{0}) = 0$, and its partial Jacobian
$\partial_{\theta} F(\thstar, D_{0})$ is invertible. Then $F = 0$
implicitly defines $\theta$ as a unique $C^{1}$ function of $d$ on a
neighbourhood of $D_{0}$.}\par}

\medskip

\noindent The three hypotheses hold here:
$F(\theta, d) = J(\theta)\T(T(\theta) - d)$ is smooth in both arguments
because $T$ is; the FOC~\eqref{eq:foc} supplies $F(\thstar, D_{0}) = 0$;
and $\partial_{\theta} F(\thstar, D_{0}) = \Astar$
of~\eqref{eq:Astar-decomp} is invertible, since the standing assumption
that $\thstar$ is a non-degenerate local minimum makes $\Astar$
positive definite, $\Astar \succ 0$. Solving the chain-rule
identity~\eqref{eq:implicit-diff} for $\partial \thhat / \partial d$
then gives the central object of this paper,
\begin{equation}
\boxed{\;\;\frac{\partial \thhat}{\partial d}(D_0; \theta_{\mathrm{init}}) \;=\; \Astar^{-1}\, \Jstar\T.\;\;}
\label{eq:dPhi}
\end{equation}

\paragraph{Remark: rank of the theory Jacobian.}\mbox{}\\
An interesting condition separate from the invertibility of $\Astar$, 
that will be illustrated in Section~\ref{sec:toys-quad}, is that $\Jstar$ has full column rank.
Even when $\Jstar$ is rank-deficient, $\Astar$ can remain positive
definite thanks to the residual-weighted Hessian; the Jacobian
$\partial \thhat / \partial d = \Astar^{-1} \Jstar\T$ is then
well-defined but inherits the rank deficiency of $\Jstar\T$, and in
the extreme case $\Jstar = 0$ collapses to zero.

\medskip

\noindent With $\theta_{\mathrm{init}}$ fixed throughout, we suppress the
second argument from this point onwards and write $\thhat(d)$ as a
shorthand for $\thhat(d; \theta_{\mathrm{init}})$.
Substituting the Jacobian~\eqref{eq:dPhi} into Eq.~\eqref{eq:sigma-mc-strategy} gives the leading-order MC
replica covariance in closed form. The next two sections work this out
in detail: Section~\ref{sec:linear} treats the linear case, where the
Taylor expansion~\eqref{eq:taylor-Phi} is exact, and
Section~\ref{sec:nonlinear} treats the general nonlinear case and
compares the result with the Bayesian posterior.

\section{Linear case: exact closed form}
\label{sec:linear}

We consider a linear theory $T(\theta) = J \theta$ with
$J \in \bbR^{\Ndat \times \Npar}$ of full column rank. The Jacobian is then
the constant matrix $J(\theta) \equiv J$,
$H \equiv 0$, $A \equiv J\T J$, and the first-order
condition~\eqref{eq:foc} has the closed-form solution
\begin{equation}
\thhat(d) \;=\; (J\T J)^{-1} J\T d.
\label{eq:linear-Phi}
\end{equation}
Because $\thhat(\cdot)$ is affine, the pushforward of any Gaussian under $\thhat(\cdot)$ is
exactly Gaussian. Writing $D_0^{(k)} = D_{0} + \eta^{(k)}$,
\begin{equation}
\thhat^{(k)} \;=\; \thhat(D_{0} + \eta^{(k)})
\;=\; \underbrace{(J\T J)^{-1} J\T D_0}_{= \, \thstar} \;+\; (J\T J)^{-1} J\T\, \eta^{(k)}.
\label{eq:linear-replica}
\end{equation}
Taking $\eta^{(k)} \sim \cN(0, C)$ gives the exact MC
replica distribution,
\begin{equation}
\boxed{\;\;\thhat^{(k)} \,\big|\, D_0 \;\sim\; \cN\!\Bigl( \thstar, \; (J\T J)^{-1} J\T C \, J (J\T J)^{-1} \Bigr).\;\;}
\label{eq:linear-result}
\end{equation}
For the simplest case $C = \sigma^{2} I$,
\begin{equation}
\thhat^{(k)} \,\big|\, D_0 \;\sim\; \cN\bigl( \thstar, \; \sigma^{2} (J\T J)^{-1} \bigr),
\label{eq:linear-iid}
\end{equation}
which is the standard ordinary-least-squares covariance.

The result \eqref{eq:linear-result} is consistent with the linear-case
expression derived from a Bayesian perspective in
Ref.~\cite{DelDebbio:2021whr} and with the MC replica distribution of
Ref.~\cite{Costantini:2024wby}.

\section{Nonlinear case: leading-order Gaussian approximation}
\label{sec:nonlinear}

For a general nonlinear theory $T$ the map $\thhat(\cdot)$ is not affine, and
the pushforward of a Gaussian under $\thhat(\cdot)$ is no longer exactly
Gaussian. The leading-order statement follows from combining the
Taylor expansion~\eqref{eq:taylor-Phi} with the Jacobian
formula~\eqref{eq:dPhi}: each replica obeys
\begin{equation}
\thhat^{(k)} \;=\; \thstar \;+\; \Astar^{-1} \Jstar\T\, \eta^{(k)} \;+\; O(\| \eta^{(k)} \|^{2}),
\label{eq:taylor1}
\end{equation}
with $\Jstar$, $\Astar$ and the fit residuals $r_{a}$ as
in~\eqref{eq:Astar-decomp}. Truncating the $O(\| \eta \|^{2})$ term
gives the leading-order MC replica distribution,
\begin{equation}
\boxed{\;\;\thhat^{(k)} \,\big|\, D_0 \;\stackrel{\text{leading}}{\sim}\; \cN\!\Bigl( \thstar, \; \Astar^{-1}\, \Jstar\T\, C \, \Jstar \, \Astar^{-1} \Bigr).\;\;}
\label{eq:nonlinear-leading}
\end{equation}

It is interesting to note that for small residuals, or $T$ close to
linear, the residual-weighted Hessian is negligible and $\Astar \approx \Jstar\T \Jstar$.
Therefore Eq.~\eqref{eq:nonlinear-leading} reduces to the linear
formula~\eqref{eq:linear-result} with $J \leftrightarrow \Jstar$.

\subsection*{Comparison with the Bayesian posterior}
\label{sec:bayes-comparison}

Equation~\eqref{eq:nonlinear-leading} gives the Gaussian approximation
of the distribution that the MC replica procedure produces. However,
it is not yet clear whether this law should be interpreted as a
faithful posterior distribution for $\theta$.
In order to test this we can compare it, as it was done in Ref.~\cite{Costantini:2024wby}, 
with the posterior distribution derived from Bayes' theorem.
In particular, for nonlinear $T$ this posterior is in
general non-Gaussian, so a comparison at the level of the full
posterior would mix orders in the small-noise expansion; to stay at
the same order as the MC result we compare instead with the
\emph{Laplace approximation} of the Bayes posterior, namely, the Gaussian
obtained by a second-order Taylor expansion of $-\log p(\theta \mid D_0)$
around its maximum a posteriori (MAP).
Let us suppose again that the experimental data are generated as
$D \sim \cN\bigl( T(\theta), \sigma^{2} I \bigr)$. The likelihood
of $D$ given $\theta$ is then
\begin{equation}
\begin{split}
p(D \mid \theta)
\;&=\; (2\pi\sigma^{2})^{-\Ndat/2}\, \exp\!\left( -\frac{1}{2\sigma^{2}}\, \| D - T(\theta) \|^{2} \right) \\
\;&=\; (2\pi\sigma^{2})^{-\Ndat/2}\, \exp\!\left( -\frac{L(\theta; D)}{\sigma^{2}} \right) .
\end{split}
\label{eq:likelihood}
\end{equation}
Under a flat (improper) prior $p(\theta) \propto 1$ on $\theta \in \bbR^{\Npar}$, Bayes' theorem gives
the posterior
\begin{equation}
p(\theta \mid D_0) \;\propto\; p(D_0 \mid \theta)\, p(\theta) \;\propto\; \exp\!\left( -\frac{L(\theta; D_0)}{\sigma^{2}} \right),
\label{eq:posterior}
\end{equation}
so the negative log-posterior is, up to a $\theta$-independent constant,
\begin{equation}
-\log p(\theta \mid D_0) \;=\; \frac{L(\theta; D_0)}{\sigma^{2}} \;+\; \mathrm{const}.
\label{eq:neglogpost}
\end{equation}
Minimising~\eqref{eq:neglogpost} is equivalent to minimising $L(\theta; D_0)$,
so the MAP estimator coincides with the OLS MLE,
$\hat\theta_{\mathrm{MAP}} = \thstar$.

The Laplace approximation is the second-order Taylor expansion of
$-\log p(\theta \mid D_0)$ around its minimiser $\thstar$. The linear
term vanishes by the FOC at $\thstar$, and the Hessian is read off
from~\eqref{eq:neglogpost} as
$\sigma^{-2}\nabla^{2}_{\theta} L(\thstar; D_0) = \Astar/\sigma^{2}$, i.e. the matrix~\eqref{eq:hessian} at $(\thstar, D_{0})$ rescaled by
$1/\sigma^{2}$. The quadratic approximation therefore reads
\begin{equation}
-\log p(\theta \mid D_0) \;=\; \mathrm{const} \;+\; \tfrac{1}{2\sigma^{2}}\,(\theta - \thstar)\T \Astar\, (\theta - \thstar) \;+\; O\bigl( \|\theta-\thstar\|^{3} \bigr),
\label{eq:laplace-expand}
\end{equation}
and exponentiating gives a Gaussian centred at $\thstar$ with
covariance $\sigma^{2}\Astar^{-1}$:
\begin{equation}
\theta \,\big|\, D_0 \;\stackrel{\text{Laplace}}{\sim}\; \cN\!\bigl( \thstar, \; \sigma^{2}\, \Astar^{-1} \bigr).
\label{eq:bayes-laplace}
\end{equation}

Comparing the MC replica covariance \eqref{eq:nonlinear-leading} with
$C = \sigma^{2} I$,
\begin{equation}
\begin{gathered}
\Sigma_{\mathrm{MC}} \;=\; \sigma^{2}\, \Astar^{-1} \Jstar\T \Jstar \Astar^{-1}, \\
\Sigma_{\mathrm{Bayes}} \;=\; \sigma^{2}\, \Astar^{-1},
\end{gathered}
\end{equation}
gives the discrepancy
\begin{equation}
\begin{split}
\Sigma_{\mathrm{MC}} - \Sigma_{\mathrm{Bayes}}
\;&=\; \sigma^{2}\, \Astar^{-1}\!\bigl( \Jstar\T \Jstar - \Astar \bigr) \Astar^{-1} \\
\;&=\; -\sigma^{2}\, \Astar^{-1}\!\left( \sum_{a=1}^{\Ndat} r_{a}\, H_{a,\cdot\cdot}(\thstar) \right)\!\Astar^{-1}.
\end{split}
\label{eq:discrepancy}
\end{equation}
The leading-order MC and Bayesian covariances differ by a term proportional
to the residual-weighted Hessian.
Three regimes are interesting to characterise.

\begin{itemize}
\item \emph{Linear theory.} $H \equiv 0$, the discrepancy~\eqref{eq:discrepancy}
vanishes identically, and the MC distribution coincides with the Bayesian
posterior exactly, by~\eqref{eq:linear-result}.
This is the well-known agreement between the two frameworks in the linear
case~\cite{DelDebbio:2021whr, Costantini:2024wby}.

\item \emph{Nonlinear theory, good fit.} If the residuals $r_{a}$ are small
in size compared to the curvature of $T$, the residual-weighted
Hessian is small and the two covariances approximately agree.

\item \emph{Nonlinear theory, poor fit.} If the residuals are not
small \emph{and} $T$ has appreciable curvature, the MC and Bayesian
covariances differ at leading order. Whether the MC procedure over-
or under-covers Bayes in a given parameter direction depends on the
sign of the residual-weighted Hessian $\sum_{a} r_{a} H_{a,\cdot\cdot}(\thstar)$
along that direction, and is studied case by case in
Section~\ref{sec:toys}.

\end{itemize}

The leading-order MC replica distribution is therefore a faithful
approximation to the Bayesian posterior whenever the residual-weighted
Hessian is negligible. When the
residual-weighted Hessian is non-negligible it is possible to use Eq.~\eqref{eq:discrepancy}
to quantify how far the MC distribution departs from the Bayesian
posterior. The next section makes these regimes concrete on two
simple toy examples, where the residuals, the curvature of the theory,
and the resulting MC--Bayes discrepancy can all be computed in closed
form.

\section{Toy examples}
\label{sec:toys}

To make the leading-order MC distribution~\eqref{eq:nonlinear-leading}
and its discrepancy~\eqref{eq:discrepancy} with the Bayes--Laplace
posterior concrete, we work out two single-parameter toy examples taken
from Ref.~\cite{Costantini:2024wby}. 

\subsection{Quadratic theory at a single data point}
\label{sec:toys-quad}

\begin{figure}[t]
\centering
\begin{tikzpicture}[scale=1.3]
  \draw[->] (-2.8, 0) -- (3.0, 0) node[right] {$\theta$};
  \draw[->] (0, -2.3) -- (0, 5.0) node[above] {$T(\theta)$};

  \draw[very thick, darkblue, domain=-2.25:2.25, samples=80] plot (\x, {\x*\x});

  \fill (0, 0) circle (1.6pt);
  \node[anchor=north west, inner sep=2pt] at (0.08, -0.05) {\small$(\theta_{\min},\, t_{\min})$};

  \draw[green!55!black, very thick, dashed] (-2.8, 4) -- (2.5, 4);
  \node[green!45!black, anchor=west] at (2.55, 4) {\small (a) $D_0>t_{\min}$};
  \fill[green!55!black] (-2, 4) circle (1.6pt) node[below left, inner sep=1pt] {\small$\theta_{-}$};
  \fill[green!55!black] ( 2, 4) circle (1.6pt) node[below right, inner sep=1pt] {\small$\theta_{+}$};
  \draw[green!55!black, very thin, dotted] (-2, 4) -- (-2, 0);
  \draw[green!55!black, very thin, dotted] ( 2, 4) -- ( 2, 0);

  \draw[orange!90!black, very thick, dashed] (-2.8, 0) -- (-0.9, 0);
  \draw[orange!90!black, very thick, dashed] (0.9, 0) -- (2.5, 0);
  \node[orange!90!black, anchor=south west, inner sep=2pt] at (2.55, 0.05) {\small (b) $D_0=t_{\min}$};

  \draw[red!80!black, very thick, dashed] (-2.8, -1.6) -- (2.5, -1.6);
  \node[red!80!black, anchor=west] at (2.55, -1.6) {\small (c) $D_0<t_{\min}$};
  \draw[red!80!black, very thick, ->] (0, -1.55) -- (0, -0.18);
  \node[red!80!black, anchor=west, inner sep=2pt] at (0.1, -0.85) {\small projection $\to \theta_{\min}$};

\end{tikzpicture}
\caption{The three data regimes for the quadratic theory~\eqref{eq:quad-T},
drawn in the canonical form $T(\theta) = \theta^{2}$ (so $\theta_{\min} = 0$,
$t_{\min} = 0$). \emph{(a) Reachable}: the line $T = D_0$ cuts the parabola
twice, at the two best-fit branches $\theta_{\pm}$
of~\eqref{eq:quad-branches}, with zero residual $r = 0$ at either branch
($\Jstar \ne 0$, $\Astar = \Jstar^{2}$). \emph{(b) Marginal}: the line $T = D_0$
is tangent to the parabola at the vertex $\theta_{\min}$, where the two
branches merge; here $J(\theta_{\min}) = 0$ \emph{and} $r = 0$ simultaneously,
so $\Astar = 0$ and our framework does not apply. \emph{(c) Unreachable}: the line
$T = D_0$ lies strictly below the parabola, so no $\theta$ achieves $T(\theta)
= D_0$; the loss $L(\theta; D_0)$ is minimised at the closest point of the theory, 
namely $\theta_{\min}$, with strictly positive residual
$r = t_{\min} - D_{0} > 0$ and Hessian $\Astar = r H > 0$. The
$\delta$-limit~\eqref{eq:quad-delta} of the framework arises only in
regime (c).}
\label{fig:quad-regimes}
\end{figure}
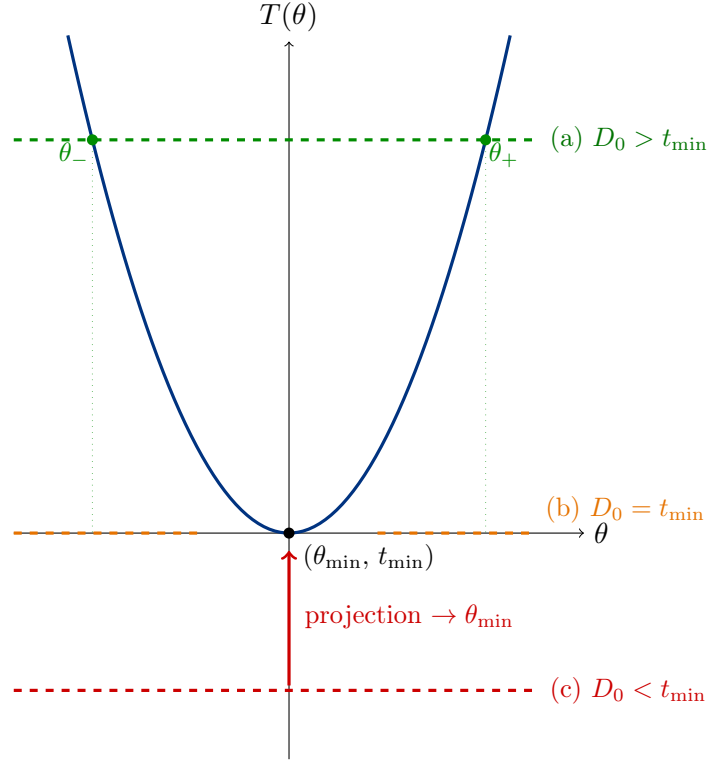

Take $\Ndat = \Npar = 1$, $\theta \in \bbR$ for the single parameter,
$D_0 \in \bbR$ for the single data point, $C = \sigma^{2}$, and
\begin{equation}
T(\theta) \;=\; t_{0} \;+\; t_{\mathrm{lin}}\, \theta \;+\; t_{\mathrm{quad}}\, \theta^{2},
\qquad t_{\mathrm{quad}} > 0,
\label{eq:quad-T}
\end{equation}
so that
\begin{equation}
J(\theta) \;=\; t_{\mathrm{lin}} \;+\; 2\, t_{\mathrm{quad}}\, \theta,
\qquad H(\theta) \;=\; 2\, t_{\mathrm{quad}}.
\label{eq:quad-derivatives}
\end{equation}
We observe that the Jacobian vanishes at the minimum of $T$,
\begin{equation}
\theta_{\min} \;:=\; -\,\frac{t_{\mathrm{lin}}}{2\, t_{\mathrm{quad}}},
\qquad
t_{\min} \;:=\; T(\theta_{\min}) \;=\; t_{0} \;-\; \frac{t_{\mathrm{lin}}^{2}}{4\, t_{\mathrm{quad}}},
\label{eq:quad-critical}
\end{equation}
where $t_{\min}$ is the minimum value of $T$ over $\theta \in \bbR$. The first-order condition~\eqref{eq:foc} reduces
in this scalar setting to
\begin{equation}
J(\theta)\, \bigl( T(\theta) - D_0 \bigr) \;=\; 0,
\label{eq:quad-foc}
\end{equation}
and is solved either by $\theta = \theta_{\min}$ (zeroing $J$) or by
$T(\theta) = D_0$ (zeroing the residual). The latter has real solutions only
for $D_{0} \ge t_{\min}$, namely the two branches
\begin{equation}
\theta_{\pm} \;=\; \theta_{\min} \;\pm\; \sqrt{(D_{0} - t_{\min}) / t_{\mathrm{quad}}}.
\label{eq:quad-branches}
\end{equation}
Three regimes follow, distinguished by the location of $D_{0}$ relative to
$t_{\min}$, as illustrated in Figure~\ref{fig:quad-regimes}.

\paragraph{(a) Reachable data, $D_{0} > t_{\min}$.}
The best fit is $\thstar \in \{\theta_{+}, \theta_{-}\}$, with one branch
selected by the initialisation convention of Section~\ref{sec:mle-map}.
At either branch $T(\thstar) = D_{0}$ exactly, so the residual is
$r := T(\thstar) - D_{0} = 0$, the residual-weighted Hessian
in~\eqref{eq:Astar-decomp} vanishes, and
\begin{equation}
\begin{gathered}
\Jstar \;\ne\; 0,
\qquad
\Astar \;=\; \Jstar^{2}, \\
\Sigma_{\mathrm{MC}} \;=\; \frac{\sigma^{2}\, \Jstar^{2}}{\Astar^{2}} \;=\; \frac{\sigma^{2}}{\Jstar^{2}},
\qquad
\Sigma_{\mathrm{Bayes}} \;=\; \frac{\sigma^{2}}{\Astar} \;=\; \frac{\sigma^{2}}{\Jstar^{2}}.
\end{gathered}
\label{eq:quad-reachable}
\end{equation}
The leading-order MC and Bayes covariances coincide in our leading order picture since
the zero residual removes the only term by which they can differ.

\paragraph{(b) Marginal data, $D_{0} = t_{\min}$.}
The two reachable branches~\eqref{eq:quad-branches} merge at
$\theta_{\min}$ and the FOC~\eqref{eq:quad-foc} has the degenerate
solution $\thstar = \theta_{\min}$ with $\Jstar = 0$ and $r = 0$, so
$\Astar = 0$. The Taylor expansion~\eqref{eq:taylor1} is undefined and our 
framework, as well as the Laplace approximation, do not apply here.

\paragraph{(c) Unreachable data, $D_{0} < t_{\min}$.}
No real $\theta$ satisfies $T(\theta) = D_0$. The unique minimiser of
$L(\theta; D_0)$ is then the closest point of the theory to $D_{0}$,
namely $\thstar = \theta_{\min}$, with $\Jstar = 0$ and
\begin{equation}
r \;=\; T(\theta_{\min}) - D_0 \;=\; t_{\min} - D_{0} \;>\; 0.
\end{equation}
Plugging into~\eqref{eq:Astar-decomp},
\begin{equation}
\Astar \;=\; \underbrace{\Jstar^{2}}_{=\,0} \;+\; r\, H \;=\; 2\, t_{\mathrm{quad}}\, \bigl( t_{\min} - D_{0} \bigr) \;>\; 0,
\label{eq:quad-Astar-unreach}
\end{equation}
and therefore $\thhat(\cdot)$ is still differentiable at $D_{0}$; only its Jacobian collapses. 
The leading-order MC covariance from~\eqref{eq:nonlinear-leading} is then
\begin{equation}
\Sigma_{\mathrm{MC}}
\;=\; \frac{\sigma^{2}\, \Jstar^{2}}{\Astar^{2}}
\;\xrightarrow[\Astar \,>\, 0]{\Jstar \to 0}\; 0,
\label{eq:quad-Sigma-MC-unreach}
\end{equation}
i.e.\ the MC replica law collapses to a $\delta$-function concentrated
at $\theta_{\min}$:
\begin{equation}
\boxed{\;\;\thhat^{(k)} \,\big|\, D_0 \;\stackrel{\mathrm{lead}}{\sim}\; \delta\!\bigl( \theta - \theta_{\min} \bigr) \qquad \text{when } D_{0} < t_{\min}.\;\;}
\label{eq:quad-delta}
\end{equation}
The Bayesian posterior in the same regime remains a well-defined Gaussian,
$\cN\bigl( \theta_{\min},\, \sigma^{2} / \Astar \bigr)$ with $\Astar$
of~\eqref{eq:quad-Astar-unreach} finite and strictly positive.

The $\delta$ in~\eqref{eq:quad-delta} is precisely the
$\delta\!\bigl( c - c^{\star} \bigr)$ contribution exhibited explicitly
in eq.~(2.29) of Ref.~\cite{Costantini:2024wby}. In the present
derivation its origin is transparent: the collapse is driven entirely
by $\Jstar \to 0$ in the Jacobian
$\partial \thhat / \partial d = \Astar^{-1}\, \Jstar\T$
of~\eqref{eq:dPhi}, with $\Astar$ kept strictly positive by the
residual-weighted Hessian alone. This is the sharpest realisation of
the rank-deficiency remark of Section~\ref{sec:mle-map}: $\thhat(\cdot)$
is still differentiable at $D_{0}$, yet the pushforward~\eqref{eq:taylor1}
of the replica noise concentrates on a single point. The Bayes--Laplace
posterior $\cN\!\bigl( \theta_{\min},\, \sigma^{2}/\Astar \bigr)$, by
contrast, remains a non-degenerate Gaussian throughout the unreachable
regime, so the qualitative gap between the MC and Bayesian
descriptions of $\theta$ is most extreme precisely here.

\subsection{Circular theory}
\label{sec:toys-circular}

Take $\Ndat = 2$, $\Npar = 1$, $\theta \in \bbR$, $D_0 = (D_{0,1}, D_{0,2}) \in \bbR^{2}$,
$C = \sigma^{2}\, I_{2}$, and
\begin{equation}
T(\theta) \;=\; t_{0} \begin{pmatrix} \cos \theta \\ \sin \theta \end{pmatrix},
\qquad t_{0} > 0.
\label{eq:circ-T}
\end{equation}
Direct differentiation gives
\begin{equation}
\begin{gathered}
J(\theta) \;=\; t_{0} \begin{pmatrix} -\sin \theta \\ \phantom{-}\cos \theta \end{pmatrix},
\qquad
J(\theta)\T J(\theta) \;=\; t_{0}^{2}, \\
H_{a}(\theta) \;=\; -\, T_{a}(\theta), \quad a = 1, 2.
\end{gathered}
\label{eq:circ-derivatives}
\end{equation}
The Gauss--Newton scalar $J\T J = t_{0}^{2}$ is a positive constant; in
particular $J(\theta)$ has full column rank for every $\theta$.
Moreover, as we show below, $\Astar \succ 0$ and therefore $\thhat(\cdot)$
is differentiable at any best fit $\thstar$.
The non-injectivity $T(\theta) = T(\theta + 2\pi k)$ of~\eqref{eq:circ-T} is not a problem since
the initialisation convention of Section~\ref{sec:mle-map} confines $\thhat(\cdot)$ to a single branch by
definition.

For $D_{0} \ne 0$, the stationarity condition $J(\theta)\T\bigl(T(\theta) - D_0\bigr) = 0$
reduces to $\sin\theta\, D_{0,1} - \cos\theta\, D_{0,2} = 0$, whose minimising solution
is $\thstar = \arg(D_{0})$; the corresponding theory point $T(\thstar) = t_{0}\, D_0 / |D_0|$
is the orthogonal projection of $D_{0}$ onto the theory circle. The residual is
\begin{equation}
r \;=\; T(\thstar) - D_{0} \;=\; \frac{t_{0} - |D_0|}{|D_0|}\, D_0,
\label{eq:circ-residual}
\end{equation}
and the residual-weighted Hessian evaluates to
\begin{equation}
\begin{split}
\sum_{a=1}^{2} r_{a}\, H_{a}(\thstar)
\;&=\; -\, r\T\, T(\thstar) \\
\;&=\; -\, \frac{t_{0} - |D_0|}{|D_0|}\, D_0\T\, \frac{t_{0}\, D_0}{|D_0|} \\
\;&=\; t_{0}\,\bigl( |D_0| - t_{0} \bigr),
\end{split}
\label{eq:circ-rH}
\end{equation}
where the first equality used $H_{a} = -T_{a}$ from~\eqref{eq:circ-derivatives}.
Combining~\eqref{eq:circ-derivatives} and~\eqref{eq:circ-rH},
\begin{equation}
\Astar \;=\; t_{0}^{2} \;+\; t_{0}\,\bigl( |D_0| - t_{0} \bigr) \;=\; t_{0}\, |D_0|,
\label{eq:circ-Astar}
\end{equation}
and is strictly positive ($\Astar \succ 0$). Substituting into~\eqref{eq:nonlinear-leading}
and~\eqref{eq:bayes-laplace},
\begin{equation}
\begin{gathered}
\Sigma_{\mathrm{MC}}
\;=\; \frac{\sigma^{2}\, t_{0}^{2}}{(t_{0}\, |D_0|)^{2}}
\;=\; \frac{\sigma^{2}}{|D_0|^{2}}, \\
\Sigma_{\mathrm{Bayes}}
\;=\; \frac{\sigma^{2}}{\Astar}
\;=\; \frac{\sigma^{2}}{t_{0}\, |D_0|},
\end{gathered}
\label{eq:circ-covariances}
\end{equation}
with leading-order discrepancy
\begin{equation}
\Sigma_{\mathrm{MC}} - \Sigma_{\mathrm{Bayes}}
\;=\; \frac{\sigma^{2}\,\bigl( t_{0} - |D_0| \bigr)}{t_{0}\, |D_0|^{2}}.
\label{eq:circ-discrepancy}
\end{equation}
Three regimes follow from the sign of $t_{0} - |D_0|$:
\begin{itemize}
\item $|D_0| < t_{0}$ (data inside the theory circle): the discrepancy
is positive,
$\Sigma_{\mathrm{MC}} > \Sigma_{\mathrm{Bayes}}$,
and the MC procedure over-covers $\theta$ relative to Bayes.
\item $|D_0| > t_{0}$ (data outside the theory circle): the discrepancy
is negative, MC under-covers.
\item $|D_0| = t_{0}$ (data exactly on the theory circle): the residual
$r$ in~\eqref{eq:circ-residual} vanishes, the residual-weighted
Hessian~\eqref{eq:circ-rH} vanishes, $\Astar = \Jstar\T \Jstar = t_{0}^{2}$,
and the two covariances coincide --- consistent with the general
``good fit'' regime of Section~\ref{sec:bayes-comparison}.
\end{itemize}
In particular, we note that our leading-order Gaussian approximation
reproduces the over- and under-coverage as it was already presented in
Fig.~2.4 of Ref.~\cite{Costantini:2024wby};
the non-Gaussian $\mathrm{erfc}$ tail of eq.~(2.36) is a higher-order
correction beyond the linear Taylor expansion~\eqref{eq:taylor1} of
$\thhat(\cdot)$, arising from its $O(\|\eta\|^{2})$ remainder.

\section{Conclusions}
\label{sec:conclusions}

We have proposed a derivation of the Monte Carlo replica distribution
that views it as the pushforward of the Gaussian replica noise
$\eta \sim \cN(0, C)$ under the MLE map $\thhat : \bbR^{\Ndat} \to \bbR^{\Npar}$,
$d \mapsto \arg\min_{\theta} \tfrac{1}{2}\| d - T(\theta) \|^{2}$.
A first-order Taylor expansion of $\thhat(\cdot)$ around $D_{0}$
reduces this pushforward to a linear-Gaussian map, whose covariance
is fixed by the Jacobian of $\thhat$ at $D_{0}$; implicit
differentiation of the first-order optimality condition then supplies
this Jacobian in closed form~\eqref{eq:dPhi}, delivering the
leading-order MC replica distribution.

For a linear theory $T(\theta) = J\theta$ the Taylor expansion is
exact: the MC distribution is exactly Gaussian and coincides with
the Bayes--Laplace posterior~\eqref{eq:linear-result}, in agreement
with the well-known equivalence of the two frameworks in the linear
case.

For a general nonlinear theory the leading-order MC
covariance~\eqref{eq:nonlinear-leading} and the Bayes--Laplace
covariance differ by a single matrix, the
residual-weighted Hessian $\sum_{a} r_{a}\, H_{a,\cdot\cdot}(\thstar)$
of the theory map at the best fit~\eqref{eq:discrepancy}. This is the
central diagnostic of the present derivation: it is built from the
residuals, the theory Jacobian $\Jstar$ and the theory Hessian
$H(\thstar)$ at the best fit, vanishes whenever the residuals are
small or the theory is locally linear, and otherwise sets the sign and
the size of the MC over- or under-coverage of the Bayesian posterior.

The two single-parameter toy examples of Section~\ref{sec:toys}
illustrate this diagnostic in closed form and, in both cases, the
leading-order picture is in agreement with the full MC replica
distribution already presented in Ref.~\cite{Costantini:2024wby}: the
quadratic theory (Section~\ref{sec:toys-quad}) reproduces
the delta-like collapse~\eqref{eq:quad-delta} of the MC distribution in the
unreachable regime, and the circular theory
(Section~\ref{sec:toys-circular}) reproduces at leading order the
over-/under-coverage reported in Fig.~2.4 of the same reference.

The discrepancy~\eqref{eq:discrepancy} provides a practical, computable
indicator that can be evaluated in realistic PDF and SMEFT fits.
This can help understand when and whether the MC replica method is a viable alternative 
to a full Bayesian analysis especially in regimes in which the Bayesian posterior 
is computationally out of reach.
A natural extension of the present analysis is to evaluate the                                                                            
residual-weighted Hessian $\sum_{a} r_{a}\, H_{a,\cdot\cdot}(\thstar)$                                                                    
in a full NNPDF PDF determination: its magnitude
quantifies how far the MC replica distribution
departs from the Bayes--Laplace posterior, and its sign along each
data point reveals whether the MC procedure over- or
under-covers Bayes for that data point.

\acknowledgments
We thank Ella Cole, Luigi Del Debbio, Stefano Forte and Maria Ubiali for their critical reading of the text and for useful
discussions. Mark N. Costantini is supported by the European Research Council
under the European Union's Horizon 2020 research and innovation
Programme (grant agreement n.950246).

\bibliographystyle{JHEP}
\bibliography{references}

\end{document}